# Distance and age of the massive stellar cluster Westerlund 1. II. The eclipsing binary W36

Danilo F. Rocha[1,2]⋆, Leonardo A. Almeida[3,2]†, Augusto Damineli[4], Felipe Navarete[5], Michael Abdul-Masih[6,7], Gregory N. Mace[8]

[1] *Observatório Nacional, R. Gen. José Cristino, 77 - Vasco da Gama, Rio de Janeiro - RJ, 20921-400, Brazil*
[2] *Programa de Pós-graduação em Física, Universidade do Estado do Rio Grande do Norte, Mossoró - RN, 59610-210, Brazil*
[3] *Escola de Ciências e Tecnologia, Universidade Federal do Rio Grande do Norte, Natal - RN, 59072-970, Brazil*
[4] *Universidade de São Paulo, Instituto de Astronomia, Geofísica e Ciências Atmosféricas, Rua do Matão, 1226, São Paulo - SP, 05508-090, Brazil*
[5] *SOAR Telescope/NSF's NOIRLab, Avda Juan Cisternas 1500, 1700000, La Serena, Chile*
[6] *European Southern Observatory, Alonso de Cordova 3107, Vitacura, Casilla 19001, Santiago de Chile, Chile*
[7] *Institute of Astrophysics, KULeuven, Celestijnenlaan 200 D, 3001 Leuven, Belgium*
[8] *The University of Texas at Austin, Department of Astronomy, 2515 Speedway, Stop C1400, Austin, TX 78712-1205*

**ABSTRACT**

Westerlund 1 (Wd 1) is one of the most relevant star clusters in the Milky Way to study massive star formation, although it is still poorly known. Here, we used photometric and spectroscopic data to model the eclipsing binary W36, showing that its spectral type is O6.5 III + O9.5 IV, hotter and more luminous than thought before. Its distance $d_{W36} = 4.03 \pm 0.25$ kpc agrees, within the errors, with three recent Gaia-EDR3-based distances reported in Paper I, Beasor & Davies, and by Negueruela's group. However, they follow different approaches to fix the zero-points for red sources such as those in Wd 1 and to select the best approach, we used an accurate modelling of W36. The weighted mean distance of our parallax (Paper I) and binary distances results in $d_{wd1} = 4.05 \pm 0.20$ kpc, with an unprecedented accuracy of 5%. We adopted isochrones based on the Geneva code with supersolar abundances to infer the age of W36B as $6.4 \pm 0.7$ Myr. This object seems to be part of the prolific star formation burst represented by OB giants and supergiants that occurred at $7.1 \pm 0.5$ Myr ago, which coincides with the recently published PMS isochrone with age 7.2 Myr. Other BA-type luminous evolved stars and Yellow Hypergiants spread in the age range of 8–11 Myr. The four Red Supergiants discussed in paper I represent the oldest population of the cluster with an age of $10.7 \pm 1$ Myr. The multiple episodes of star formation in Wd 1 are reminiscent of that reported for the R136/30 Dor LMC cluster.

## 1 INTRODUCTION

Westerlund 1 (Wd 1) represents one of the most massive and dense star clusters in our Galaxy, with approximately $10^5 M_\odot$ (Clark et al. 2005). It contains a variety of isolated and binary sources classified as OB Giants and Supergiants (OBSGs), Red Supergiants (RSGs), Yellow Hypergiants (YHGs), and Wolf Rayet Stars (WRs) at the final evolutionary stages of high-mass stars (Bonanos 2007, hereafter B07). Previous works have reported distances between 2-6 kpc (Clark et al. 2005; Kothes & Dougherty 2007; Andersen et al. 2017; Davies & Beasor 2019; Beasor et al. 2021) to Wd 1, but the uncertainties involved are relatively large, translating into a broad range of values for the physical parameters of the cluster and its members (e.g., ages and masses). For this reason, the characterisation of relatively distant objects requires accurate estimates of both the distance and the amount of light lost due to local and interstellar medium (ISM) extinction.

To date, the reddening law and the mean $A_{K_s}$ extinction are the best well-constrained parameters for the Wd 1 cluster. The extinction towards Wd 1 was investigated by Damineli et al. (2016, hereafter D16), who derived an accurate reddening law for the inner Galactic plane. Those authors obtained an average extinction of the spectroscopically classified stars as $A_{K_s} = 0.74 \pm 0.01$ mag. However, the mean $A_{K_s}$ value is limited to a sample of spectroscopically classified stars from Clark et al. (2020), which is biased to lower extinction and brighter members, as pointed out by Negueruela et al. (2022). However, it seems that the extinction is not much larger, since Hosek et al. (2018) reported $A_{K_s} = 0.78 \pm 0.18$ mag, using a brightness-unbiased sample.

The distance of Wd 1 has been reported by several authors using different methods over the years. Following Clark et al. (2005), the lower and upper limits for Wd 1 distance would be ∼2 kpc (inferred via near-infrared colours of 18 WR stars) and ∼5.5 kpc (from yellow supergiants luminosities), respectively. Other distance estimates were reported from radial velocity (RV) studies of an H I feature (∼ 3.90 kpc, Kothes & Dougherty 2007), and on the observed near-infrared magnitude of the transition region between pre-main sequence and main-sequence (3.55 and 4.0 kpc, Brandner et al. 2008; Gennaro et al. 2011, respectively). Recently, using parallaxes from the Gaia Data Release 2 (Gaia DR2), Davies & Beasor (2019) derived a distance of 3.87 kpc, corroborating the distance of 3.78 kpc from Rate et al. (2020), and 4.12 kpc of Beasor et al. (2021) based on parallaxes from the Gaia Early Data Release 3 (Gaia EDR3). All these Gaia distances contrast with the Gaia DR2 result from Aghakhanloo et al. (2020), 2.60 kpc. The discrepancy between these values is a strong limitation to assessing the basic parameters of stellar populations in Wd 1 and, in particular, the age of the cluster itself.

---

⋆ E-mail: danilo-fr@live.com
† E-mail: leonardo.almeida@ufrn.br

Eclipsing binaries provide a robust and direct method to measure stellar distances up to extragalactic scales (Bonanos et al. 2006). This method has been applied to improve the calibration of the Hubble constant $H_0$ in low redshifts (e.g., Pietrzyński et al. 2013, 2019), as well as to determine the age of stellar clusters (e.g., Meibom et al. 2009; Thompson et al. 2020). In the context of Wd 1, Koumpia & Bonanos (2012) modelled four eclipsing binaries based on their optical photometry and spectroscopy, and using the W13 system derived the distance of 3.71 ± 0.55 kpc to the cluster. However, their work has two drawbacks: *i)* the adoption of the reddening law from Indebetouw et al. (2005), which is not adequate for the inner Galaxy; and *ii)* the date of the *J*-band measurement is unknown, making it impossible to know the orbital phase to recover the brightness of the objects outside the eclipses (phases 0.25 or 0.75). Moreover, the W13 light curve is not clean, especially in the *V* and *R* bands, with random scattering of ∼0.1 mag. Recently, Hosek et al. (2018) recalculated the distance to W13 as 3.90 ± 0.40 kpc, using the same photometric observations reported by Koumpia & Bonanos (2012) but adopting their updated reddening law.

Navarete et al. (2022, hereafter Paper I) fine-tuned procedures to improve the accuracy of the distance determination to the Wd 1 cluster using astrometry and photometry from the Gaia-EDR3. They used a sample of 172 stars with Gaia-EDR3 counterparts and obtained a distance of $d_{\rm gedr3} = 4.06^{+0.36}_{-0.34}$ kpc to the cluster.

To check for the reliability of the systematic effects in the Gaia-EDR3 astrometry, we further derived an independent distance by revisiting the distance with the classical method of eclipsing binaries which can give accuracy up to 10% for the Local Group of Galaxies, when using OB stars (Bonanos et al. 2006; Koumpia & Bonanos 2012). We performed a re-analysis of the W36[1] eclipsing binary reported by Koumpia & Bonanos (2012), by adding new photometry taken in the time-frame of 2006-2020 and optical and near-infrared (NIR) spectroscopy from FLAMES/VLT, IGRINS/Gemini South, and TripleSpec 4.1/SOAR, respectively.

This manuscript is organised as follows. In Sec. 2, we present the photometric and spectroscopic data used in this paper and the reduction procedure. Section 3 presents the spectral characterisation of the W36 components and the derivation of their orbital and physical parameters. Finally, we discuss the implications of our results and summarise our conclusion in Sect. 4.

## 2 DATA

### 2.1 Spectroscopic observations

We used spectroscopic observations taken at FLAMES/VLT to construct the radial velocity curve of the eclipsing binary stellar system W36. We retrieved the FLAMES/VLT reduced data from the ESO archive (Program ID 091.D-0179; PI: J.S. Clark), observed in

---

[1] RA = 16h47m05.08s, Dec = −45°50′55″.1, J2000.

**Table 1.** Summary of spectroscopic observations of W36.

| Epoch | Date | Instrument | $t_{\rm exp}$ (sec) | R ($\lambda/\Delta\lambda$) |
|---|---|---|---|---|
| 1a | 2013/07/10 | FLAMES/VLT[a] | 2445 | 18000 |
| 1b | 2013/07/10 | FLAMES/VLT[a] | 2445 | 18000 |
| 2 | 2013/07/21 | FLAMES/VLT[a] | 2445 | 18000 |
| 3 | 2013/08/08 | FLAMES/VLT[a] | 2445 | 18000 |
| 4a | 2013/08/18 | FLAMES/VLT[a] | 2445 | 18000 |
| 4b | 2013/08/18 | FLAMES/VLT[a] | 2445 | 18000 |
| 5 | 2013/09/05 | FLAMES/VLT[a] | 2445 | 18000 |
| 1 | 2020/02/03 | IGRINS/Gemini South[b] | 190 | 45000 |
| 2 | 2020/02/08 | IGRINS/Gemini South[b] | 190 | 45000 |
| 1 | 2022/05/10 | TripleSpec/SOAR[b] | 240 | 3500 |
| 2 | 2022/05/15 | TripleSpec/SOAR[b] | 240 | 3500 |
| 3 | 2022/05/16 | TripleSpec/SOAR[b] | 240 | 3500 |
| 4 | 2022/05/17 | TripleSpec/SOAR[b] | 320 | 3500 |
| 5 | 2022/06/13 | TripleSpec/SOAR[b] | 480 | 3500 |
| 6 | 2022/06/14 | TripleSpec/SOAR[b] | 480 | 3500 |
| 7 | 2022/07/30 | TripleSpec/SOAR[b] | 360 | 3500 |

**Notes:** Dates are given in YYYY/MM/DD format; *a*) ESO/Archive; *b*) this work.

2013 from July to September. Each observed epoch consists of three spectra taken with a total exposure time of 2445 sec, covering the 8481-8992 Å region with a spectral resolving power of $R \sim 18,000$. A summary of the FLAMES/VLT spectroscopic observations is presented in Table 1.

High-resolution HK spectroscopy ($R = \lambda/\Delta\lambda \sim 45,000$) of W36 was obtained using the Immersion GRating INfrared Spectrometer (IGRINS; Park et al. 2014), a long-term visiting instrument installed at the Gemini South Observatory (Mace et al. 2018), Chile (Proj. ID. GS-2020A-Q-133, PI: A. Damineli). IGRINS spectra were taken on two occasions for W36, close to phase 0.25 and 0.75. The data were reduced using the IGRINS pipeline package, which optimally extracts one-dimensional spectrum from AB nodded pairs of spectra (for more details, see Lee 2015). A summary of the IGRINS/Gemini spectroscopic observations is presented in Table 1.

Additional medium-resolution ($R \sim 3500$) NIR spectra of W36 were taken with the TripleSpec 4.1 NIR Imaging Spectrograph (Schlawin et al. 2014) at the 4.1 m SOAR telescope, Chile (PI: F. Navarete), covering the 0.9-2.5 $\mu$m region simultaneously. TripleSpec data were taken in seven dates from 2022 May to 2022 July (see Table 1), covering distinct phases of the orbital system. The data were processed using a modified version of the IDL `Spextool` package (Cushing et al. 2004; Vacca et al. 2003), providing a single one-dimensional spectrum per date. TripleSpec data was essential for obtaining a reasonable Spectral Type classification for W36 (see Sect. 3.1).

### 2.2 Photometric observations

#### 2.2.1 Optical data

Optical photometric observations of W36 were carried out using the 1.6-m Perkins-Elmer telescope at the *Observatório do Pico dos Dias* (OPD[2], Brazil). A standard set of calibration frames (bias and flat-field images) was also obtained for each night.

---

[2] OPD is operated by the *Laboratório Nacional de Astrofísica* (LNA/MCTI)

**Table 2.** Summary of photometric observations of W36.

| Date | Telescope | $t_{\rm exp}$ (sec) | Filter |
|---|---|---|---|
| 2006/06/15-07/25 | Swope-1.00m[a] | - | V |
| 2012/08/08 | LNA/OPD-1.60m[b] | 60 | I |
| 2012/08/09 | LNA/OPD-1.60m[b] | 40 | I |
| 2012/08/10 | LNA/OPD-1.60m[b] | 40 | I |
| 2012/08/13 | LNA/OPD-1.60m[b] | 20 | I |
| 2012/08/14 | LNA/OPD-1.60m[b] | 20 | I |
| 2012/08/15 | LNA/OPD-1.60m[b] | 20 | I |
| 2012/08/16 | LNA/OPD-1.60m[b] | 20 | I |
| 2019/05/20-06/17 | TESS Satellite[c] | 1426 | TESS |

**Notes:** Dates are given in YYYY-MM-DD format. *a*) B07, *b*) This work, *c*) TESS/Archive

We also used two photometric data sets: *i*) *V* data taken in 2006 with the 1-m Swope telescope from the Las Campanas Observatory (Chile), and previously reported by B07; and *ii*) observations from the Transiting Exoplanet Survey Satellite (TESS, Ricker et al. 2015), collected in 2019. A summary of the optical photometric observations is listed in Table 2.

The processing of the OPD optical photometric data was performed using IRAF[3] tasks. A master bias frame was subtracted from each science image, and the resulting image was divided by a normalised flat-field model. As our target W36 is spatially resolved in the OPD data, differential aperture photometry was adopted to extract the fluxes. We use an automatic IRAF task that executes the procedures described above and performs differential photometry.

The extraction of the TESS light curves was performed using the Python-based `Lightkurve` tool (Lightkurve Collaboration et al. 2018).

Finally, we have also included the data of the *V*-band light curve reported by B07, collected over 17 nights, between June 15 and July 25, 2006, in the Johnson system's photometric band.

### 2.2.2 Near-infrared photometry

The $JHK_s$ magnitudes of W36 were obtained at the SOAR telescope using the Spartan/SOAR camera on July 6th, 2006. The observing date corresponds to the orbital phase of 0.03 based on Eq. 2 that was derived from photometric data collected at the same epoch by Bonanos (2007). The $JHK_s$ magnitudes obtained were 10.506±0.033, 9.327±0.048, and 9.091±0.059, respectively.

$JHK_s$ photometry of OB stars is reported in Table B1. Magnitudes were extracted with PSF photometry and calibrated with 2MASS sources in the field. Statistical errors are a few millimagnitudes, but the systematic uncertainty in the calibration is 0.05 mag. The $A_{K_s}$ extinction was measured from the colour excess using the D16 reddening law. The luminosity was derived using the distance of 4.05 kpc (see Section 4). Spectral types are from Clark et al. (2020). For WRs, we used the results from Rosslowe (2015), but the luminosities were scaled to our adopted distance. Photometry of RSGs and YHGs were reported in Paper I.

---

[3] IRAF is distributed by the National Optical Astronomy Observatory, which is operated by the Association of Universities for Research in Astronomy (AURA) under cooperative agreement with the National Science Foundation.

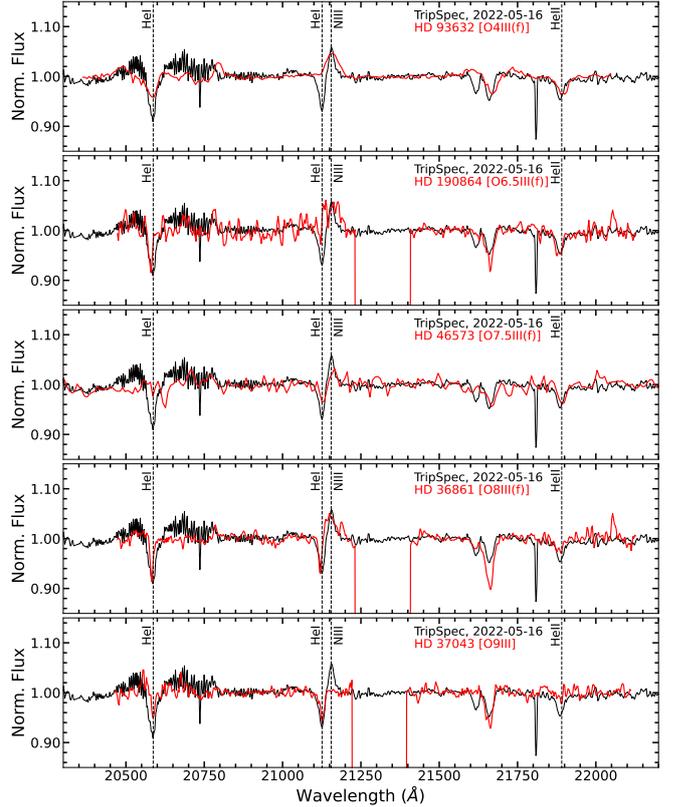

**Figure 1.** Continuum-normalised TripleSpec *K*-band spectrum of W36 used for deriving spectral type of the W36 A component. The black curves correspond to the spectrum taken on 2022-05-16 (W36 A is partially in front of W36 B, phase = 0.4). The red curves are a sample of representative O-type stars from Hanson et al. (1996). The vertical black dashed lines indicate the spectral lines used for classification of W36 A.

## 3 RESULTS

### 3.1 Spectral type classification of W36A

To proper model the binary system, it is crucial to have some prior information to constrain the parameters space (see Sect. 3.2.2). In particular, the spectral type classification provides important constraints on the effective temperature. To do that, we performed the spectral type classification of the primary component of W36 (W36 A) by comparing the normalised *K*-band spectra with the NIR spectral atlases of OB stars from Hanson et al. (1996).

Figure 1 presents the K-band spectrum of W36 taken at 2022-May-16 (in black) overlaid by spectra of stars with spectral types ranging from O4 III to O9 III from Hanson et al. (1996) which best matches the spectral features observed in the spectrum of W36. Important photospheric spectral features used for the classification are also indicated in the figure. A He I absorption at 2.0581 $\mu$m is detected within the partially corrected telluric absorption band at the blue end of the spectrum. No emission components of the C IV triplet (2.069, 2.078, and 2.083 $\mu$m), typical for spectral types from O4 to O6, are identified.

A relatively broad emission of N III is identified at 2.1155 $\mu$m, together with the He I absorption at 2.1126 $\mu$m.

The photospheric Br$\gamma$ feature exhibits a double absorption profile, arising from both A and B components. Finally, a relatively strong He II absorption is observed at 2.1885 $\mu$m.

As shown in Fig. 1, the best match for the K-band spectrum of W36 corresponds to HD 180864, an O6.5 III star exhibiting similar strengths both in the He I and He II absorption lines and the N III emission feature. As the primary star dominates the system, as seen by the depth of the eclipses (e.g. see Fig. 5), the adopted spectral type classification represents more closely the W36 A component alone.

### 3.2 Eclipsing binary modelling scheme

Eclipsing binary systems provide a straightforward method for measuring geometrical and physical parameters (e.g., effective temperature, masses, and radii) of their stellar components. Hence, they are a powerful tool for determining distances. To model the optical and near-infrared observations and to obtain the parameters for the W36 eclipsing binary, we adopted the steps described below:

(i) The fitting of the He I $\lambda$ 17002 Å coupled with He II $\lambda$ 21885 Å line profiles using the SPAMMS routine (Abdul-Masih et al. 2020) to disentangle the contribution of the primary (W36 A) and secondary (W36 B) components, and to infer their effective temperatures ($T_{1,2}$);

(ii) The fitting of the RV curves of both components to determine their semi-amplitudes ($K_1$ and $K_2$) and consequently derive the mass ratio.

(iii) With the prior information (the temperatures and mass ratio), we compared the light and the radial velocity curves with the synthetic curves generated by the Wilson-Devinney Code (WDC, Wilson & Devinney 1971) to obtain the geometrical and physical parameters of W36. To improve our search for the parameters, we simultaneously adjusted synthetic light and radial velocity curves generated by WDC implemented with the `emcee` code (Foreman-Mackey et al. 2013).

(iv) The adoption of the effective temperatures to derive the $BVRIJHK_s$ intrinsic colour indices from Wegner (1994);

(v) Use the D16 reddening law to derive the $A_{K_s}$ of W36 based on its intrinsic $JHK_s$ colours, and then, infer the de-reddened intrinsic magnitudes at $V$-band and calculate the corresponding extinction.

(vi) Use the de-reddened $V$-band light curve and the geometrical and physical parameters of W36 to derive the distance to the system.

#### 3.2.1 Ephemeris, temperatures and semi-amplitude of radial velocity of W36

We have analysed all the photometric data collected in the OPD Observatory together with the light-curves from B07 and observed by TESS in 2019, to obtain a linear ephemeris for the W36 system using the following equation,

$$T_{\min} = T_0 + P_{\text{bin}} \cdot E, \quad (1)$$

where $T_{\min}$ corresponds to the time of the minimum of the primary eclipse, $T_0$ is the initial epoch, $E$ is the number of the cycle counted from $T_0$ (in BJD units), and $P_{\text{bin}}$ is the binary orbital period (in days). The individual analysis of each photometric dataset (B07, OPD, and TESS) led to the following ephemeris:

$$T_{\min,\text{B07}} \text{ (BJD)} = 2\,453\,909.8102 + 3.1799 \cdot E, \quad (2)$$

$$T_{\min,\text{OPD}} \text{ (BJD)} = 2\,456\,149.3724 + 3.1595 \cdot E, \text{ and} \quad (3)$$

$$T_{\min,\text{TESS}} \text{ (BJD)} = 2\,458\,630.5573 + 3.1973 \cdot E, \quad (4)$$

suggesting that the system's orbital period varies over time. A detailed analysis of this variation is beyond the scope of this paper and will be covered in detail in a future work (Rocha et al. in preparation).

As the orbital period of W36 varies, we could not use all spectra available for this system to obtain the radial velocity curve. Thus, we chose to use only the FLAMES/VLT spectra taken from July 10th to September 5th, 2013 (see Table 1) to exclude any significant effect introduced by the system orbital period variations.

W36 is a double-lined spectroscopic binary (SB2). To measure the RV of the components, we followed the same procedure presented in (Sana et al. 2013; Almeida et al. 2017; Villaseñor et al. 2021). In short, two Gaussian profiles were fitted to the Pa-11 $\lambda$ 8862 Å absorption lines. We adjusted both lines simultaneously by making the depth and full width at half maximum (FWHM) of each profile to be identical throughout all the epochs. The spectra collected on August 18th, 2013 (Epoch 4b, see Fig. A1) were used as initial guesses for the Gaussian profiles as they exhibit the best signal-to-noise ratio and well-separated lines for both the components.

To identify the lines of the W36 A and W36 B components, we used the fact that the hotter primary has a shallower Pa-11 $\lambda$ 8862 Å line than the colder secondary. The Emcee code was used to search for the best solution and sample the parameter space to obtain their respective uncertainties. As the spectral lines of the two components are blended in some epochs, we could only obtain unique solutions for seven dates. The result of the fitting is presented in Fig. A1. Using these measurements, the ephemeris to the radial velocity solution is,

$$T_{\min,\text{Flames}} \text{ (BJD)} = 2\,456\,483.3719 + 3.1722 \cdot E. \quad (5)$$

The radial velocity measurement of each component in each epoch was determined from the weighted average of the 3 observed spectra and the results of both components are listed in Table 3. Fitting the radial velocity curves of the components yields semi-amplitudes $k_1 = 198^{+15}_{-10}$ km s$^{-1}$ and $k_2 = 266.2^{+6.4}_{-8.1}$ km s$^{-1}$, and systemic velocity $\gamma = -6^{+15}_{-13}$. The result of the fitting procedure is shown in Fig. 3.

To obtain an independent measurement of the temperature of W36 A and W36 B, we used the `Spectroscopic Patch Model for Massive Stars` code (SPAMMS Abdul-Masih et al. 2020) to fit the IGRINS spectra. SPAMMS is a spectral analysis tool specifically designed to model distorted massive stars to reproduce the geometry and surface variations observed in such systems. As a result, the code assigns synthetic line profiles to each patch across the surface, accounting for various local conditions such as temperature, surface gravity and radial velocity. These line profiles are then integrated across the visible surface, yielding integrated line profiles corresponding to the given phase and orientation of the system.

We used SPAMMS to model the He I $\lambda$ 17007 Å and He II $\lambda$ 21885 Å lines. We computed a grid of SPAMMS models, varying only the effective temperature of the W36 A and W36 B (from 30 000 to 40 000 K in steps of 500 K) and the Helium abundance (from 0.05 to 0.80, in steps of 0.05) in an attempt to fit the observed spectra. The best solution is found via chi-square minimisation (for more details, see Abdul-Masih et al. 2020). The best solution corresponds to the temperatures of $T^1_{\text{eff}} = 37.1^{+1.0}_{-1.1} \times 10^3$ K and $T^2_{\rm eff} = 35.4^{+0.4}_{-0.9} \times 10^3$ K and He abundance $n(He)/n(H) = 0.30 \pm 0.05$ (see Fig. 2).

**Table 3.** Radial velocity measurements of the W36 components derived from the FLAMES/VLT spectra.

| Eclipse timing BJD(+2,450,000) | Orbital phase | W36 A (km s$^{-1}$) | W36 B (km s$^{-1}$) |
|---|---|---|---|
| 6 483.7357 | 0.11 | -146.9 ± 11.8 | 184.2 ± 12.3 |
| 6 483.7714 | 0.13 | -161.0 ± 13.1 | 196.1 ± 14.0 |
| 6 494.6993 | 0.57 | 78.3 ± 16.6 | -133.0 ± 16.3 |
| 6 512.7110 | 0.25 | -179.8 ± 11.3 | 245.7 ± 16.7 |
| 6 522.5461 | 0.35 | -172.0 ± 9.2 | 212.9 ± 12.3 |
| 6 522.6599 | 0.39 | -158.1 ± 12.7 | 189.8 ± 14.5 |
| 6 540.5654 | 0.03 | -91.4 ± 12.1 | 58.9 ± 18.7 |

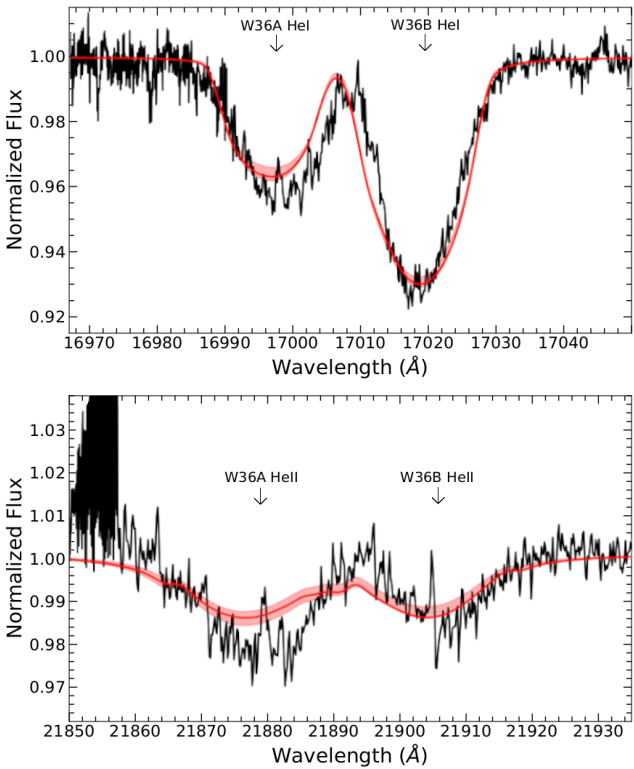

**Figure 2.** Comparison between the He I $\lambda\,17007$ Å (top panel) and He II $\lambda\,21885$ Å lines (bottom) for W36 at phase 0.67. The synthetic spectrum generated by the best-fit solution from SPAMMS are indicated by the red curves. At such orbital phase, the components of the primary and secondary stars are blue- and red-shifted, respectively. W36 A shows the faintest He I $\lambda\,17007$ Å absorption component and the relatively strongest He II $\lambda\,21885$ Å component.

### 3.2.2 Characterisation of W36

We adopted a similar procedure as described in Almeida et al. (2015) to model the W36 eclipsing binary and obtain its distance. In short, the light and radial velocity curves generated by WDC are used as a "function" to be optimised when compared with the observational data by using emcee (Foreman-Mackey et al. 2013), a Markov Chain Monte Carlo (MCMC) ensemble sampler.

This method requires a relatively well-sampled light curve in one band (e.g. *V*-band) and the radial velocity curves of the components to obtain the physical and geometrical parameters of the system (e.g., inclination, eccentricity, mass ratio, radii, masses and effective temperatures of the components). As input for the WDC and emcee, we used the radial velocity curves obtained from FLAMES/VLT spectra, and the *V*-band light curve from B07. The *V*-band data from B07 corresponds to the data associated with the slightest discrepancies between the photometric calibration presented by those authors and Clark et al. (2005).

In order to reduce and constrain the probed parameter space in our procedure, prior information obtained in Sect. 3.2.1 were used. From spectral line fitting, we kept fixed the effective temperature of the primary component ($T_1 = 37100\,K$), and from radial velocity fitting, the mass ratio (*q*) and the systemic velocity ($\gamma$) were constrained within 5 $\sigma$. Since the centres of the primary and secondary eclipses are separated by exactly 0.5 in phase, we fixed the eccentricity as zero. The orbital period and initial epoch for the light and radial velocity curves were fixed from the linear ephemeris fitting, Eqs. (2) and (5), respectively. The linear limb darkening coefficients ($\alpha$) from Claret & Bloemen (2011) were assumed for both stars. The other main parameters, e.g., (i) the orbital inclination (*i*); (ii) the effective temperature of the secondary component ($T_2$); (iii) the gravitational potentials of the primary and secondary components ($\Omega_1$ and $\Omega_2$); and (iv) the semi-major axis (*a*) was left as a free parameter in the fitting procedure.

**Table 4.** Best fit parameters from the simultaneous modelling of the *V*-band light curve and radial velocity curves for W36 performed via WDC and emcee.

| Adjusted Parameters | This work | Koumpia & Bonanos (2012) |
|---|---|---|
| $q=M_2/M_1$ | $0.71^{+0.10}_{-0.09}$ | $0.69 \pm 0.03$ |
| $\Omega_1^a$ | $3.26^{+0.18}_{-0.15}$ | $3.29 \pm 0.03$ |
| $\Omega_2^a$ | $3.36^{+0.20}_{-0.18}$ | 3.28 (fixed) |
| $T_2$ (K) | $30\,833^{+1\,554}_{-1\,539}$ | $25\,500 \pm 2\,900$ |
| $i$ (°) | $74.73^{+0.91}_{-0.76}$ | $73.0 \pm 1.7$ |
| $a^b$ (R$_\odot$) | $30.91^{+1.19}_{-1.16}$ | $27.5 \pm 1.5$ |
| $\gamma$ (km s$^{-1}$) | $-7.86^{+10.86}_{-11.15}$ | $-37 \pm 2$ |
| Derived parameters | | |
| $M_1$ (M$_\odot$) | $22.91^{+2.88}_{-2.51}$ | $16.3 \pm 1.5$ |
| $M_2$ (M$_\odot$) | $16.28^{+2.88}_{-2.31}$ | $11.3 \pm 1.8$ |
| $R_1^{\rm mean}$ (R$_\odot$) | $12.69^{+0.57}_{-0.54}$ | $11.0 \pm 1.2$ |
| $R_2^{\rm mean}$ (R$_\odot$) | $10.46^{+0.63}_{-0.67}$ | $9.2 \pm 1.2$ |
| $\log g_1$ [cm s$^{-2}$] | $3.59^{+0.02}_{-0.02}$ | $3.57 \pm 0.10$ |
| $\log g_2$ [cm s$^{-2}$] | $3.61^{+0.04}_{-0.03}$ | $3.56 \pm 0.13$ |
| $\log L_1/L_\odot$ | $5.44^{+0.04}_{-0.04}$ | $5.19 \pm 0.13$ |
| $\log L_2/L_\odot$ | $4.95^{+0.05}_{-0.06}$ | $4.51 \pm 0.23$ |

$^a$ Roche surface potential;
$^b$ Binary separation.

In this step, we choose the solution in arbitrarily scaled flux mode, and the *V*-band magnitude at phases 0.25 and 0.75 were fixed to 18.85 mag. As presented in Fig. 3, the best solution for W36 was obtained using the WDC mode 2, which puts no constraints on the Roche lobe filling. One chain with 100 000 iterations were performed via emcee to obtain the best value of the parameters and their associated error bars. Each free parameter and its error (see Table 4) were estimated by assuming the mean and the 1-$\sigma$ threshold in the range of each parameter adjusted directly by emcee. The same statistical criteria were adopted for the derived parameters, i.e. the mass, radius, and log g of each component. The adjusted temperature and luminosity of W36 B, interpolated in Martins et al. (2005) tables, results in a spectral classification O9.5 IV.

Table 4 compares ours and Koumpia & Bonanos (2012) parameters for the W36 eclipsing binary. The largest difference is in the effective temperatures, due to our NIR spectroscopy. The radii of the stars also are larger in our modelling and these two parameters have a large impact in the luminosity. These combined effects added to the fact that we use a different reddening law and extinction explains the difference in the distances obtained by these two works (see Sect. 3.2.3), which are based in the same V-band light curve reported by Bonanos (2007).

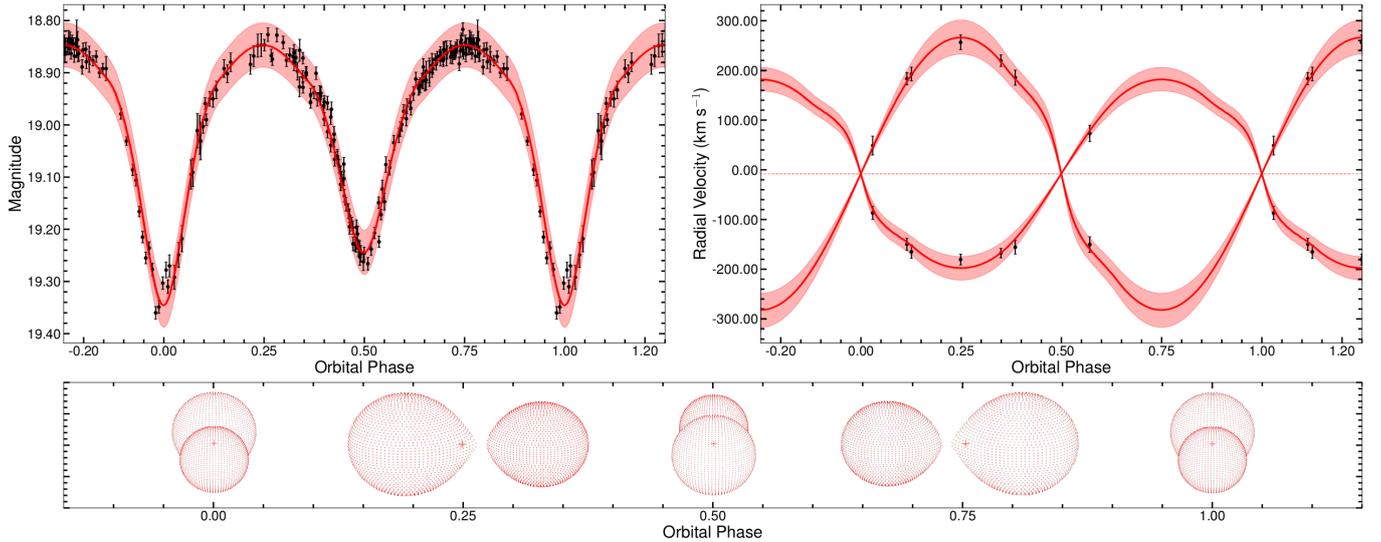

**Figure 3.** Left panel: *V*-band light curve of W36 from B07. Right panel: the radial velocity curve of W36 based on FLAMES/VLT observations. The observational data are indicated by black dots, the red-shaded areas represent the 3-$\sigma$ limits around the best fit generated by the WDC (solid red line). Bottom panel: snapshots of W36 system in 5 orbital phases. The mass centre of the system is shown with a plus sign.

*3.2.3 Extinction correction of the V−band light curve and the distance to Wd1*

The correction for extinction is commonly a source of large errors in Wd 1. To keep them as small as possible, we followed the steps described below to minimize the uncertainties introduced by the extinction correction at the *V*-band towards Wd 1.

As both W36 A and B components have similar NIR colours and the primary star is brighter than its companion (see Table 4), we used the NIR intrinsic colours of the system to infer the $E(J-K_s)$, $E(J-H)$, and $E(H-K_s)$ colour excesses. By adopting the D16 reddening law, we evaluated the total extinction at the $K_s$-band as $A_{K_s} = 0.69 \pm 0.06$ mag.

Based on the spectral classification of W36 A (O6.5 III - see Sect. 3.1) and adopting the spectral type O9.5 IV for W36 B after interpolating the values reported by Martins et al. (2005, see their Tables 4, 5, and 6) for its effective temperature and luminosity (see Table 4 ans Section 3.2.2), we derive the $(V-K_s)_0$ colour index for each component. Thereafter, the luminosities of the W36 components (see Table 4) were used as weights to derive the mean $(V-K_s)_0$ for the binary.

As we have the $K_s$ magnitude observed for the same epochs, the mean $(V-K_s)_0$ values were used to proper de-redden the *V*-band light curve. In particular, we have an accurate measurement of $K_s = 9.091 \pm 0.059$ mag on July 6th, 2006 (BJD = 2,453,922.62714), corresponding to phase = 0.03 using the B07 linear ephemeris (Eq. (2)). Thus, the corresponding $A_V$ at that phase can be obtained as:

$$A_V = V_{\text{obs},0.03} - K_{s,\text{obs},0.03} - (V-K_s)_0. \quad (6)$$

Finally, the difference in *V*-band magnitude between the phases 0.03 and 0.25 ($\Delta V_{0.03,0.25} = 0.424 \pm 0.016$) can be used to find the brightness of the system in quadrature (phases 0.25 e 0.75), i.e., outside the eclipses.

The derived extinction towards W36, $A_V = 11.725 \pm 0.099$ mag, is close to the average extinction of Wd 1 reported by D16 ($A_V = 11.26 \pm 0.07$ mag). The $A_V$ value is just a sanity check since the intrinsic $V_0$ is anchored in the $(V-K_s)_0$ colour index and the $A_{K_s}$ value, avoiding to use the reddening law for the optical range.

As the $A_{K_s}$ extinction and the colour indices have smaller errors, the uncertainties of this method are at least ten times smaller than directly applying the average reddening law.

After de-reddening the *V*-band light curve, we run again the WDC plus emcee with the same setup used in the previous section, but now, the solution in absolute flux mode (Wilson et al. 2010) in order to derive the distance to W36. The one- and two-dimensional posterior distributions of the main fitted parameters derived by the WDC and emcee are presented in Fig. 4. The resulting distance to the W36 binary system is $d_{W36} = 4.03 \pm 0.25$ kpc. The error on the distance to W36 was obtained by using the partial derivatives of the distance modulus at the *V*-band, considering the uncertainty of the apparent magnitude $\sigma_{m_V} = 0.04$ mag, the error in the extinction ($\sigma_{A_V}$, see Sect. 3.2.3), and the error in the zero point of the photometry of 0.06 mag obtained by Bonanos (2007).

To verify if our result is consistent, we use the distance module of each component to verify the value of the individual distances, and compare with the measured distance for the system. With the value of the luminosity ratio measured, the apparent magnitude in the *V*-band at phase 0.25 and the absolute magnitudes of each component correspondent to its spectral type from Martins et al. (2005), we have as a result $d_{W36\,A} = 4.06$ kpc and $d_{W36\,B} = 4.36$ kpc. The difference is 0.30 kpc, compatible with the distance error using both components ($\pm 0.25$ kpc).

The weighted mean distance of Wd 1 using the distance to W36 and the independent value obtained from the Gaia-EDR3 analysis in Paper I is $d_{\text{Wd1}} = 4.05 \pm 0.20$ kpc.

We further used our *JHKs* photometry to revise the distance of eclipsing binary W13 from Koumpia & Bonanos (2012) On JD = 2 451 318.855, which corresponds to $\phi = 0.2$, we obtained with the Spartan/SOAR camera the magnitudes $J = 9.301$, $H = 8.359$ and $Ks = 7.66$ (with extinction $A_J = 2.32$, $A_H = 1.22$, and $A_{K_s} = 0.71$). The $Ks$ measurement was close to the saturation level in order that we relay on J- and H-band only. Since at this phase the system is viewed almost in quadrature, we do not need to correct for eclipsing effects. Using the D16 reddening law, we obtain the distance of $d_{W13} = 3.75 \pm 0.07$ kpc which is compatible with values reported by Koumpia & Bonanos (2012, 3.7 ± 0.6 kpc) and also with Hosek et al.

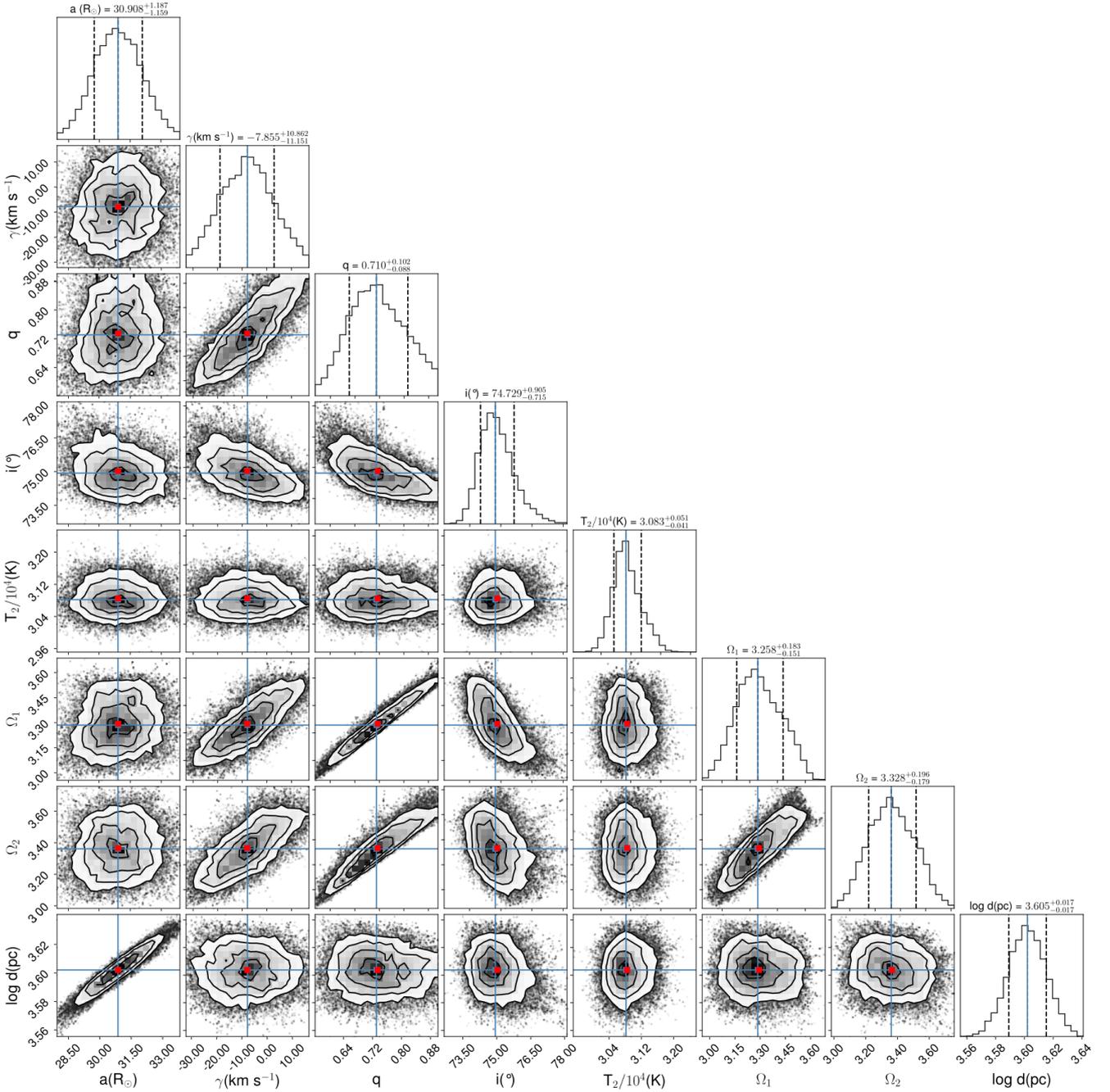

**Figure 4.** The corner plot showing the one- and two-dimensional posterior distributions for each free parameter from the simultaneous fitting of the radial velocity and *V*-band light curves of W36 performed via WCD and `emcee`. The density of the points and contours correlate with the posterior probability distribution from a total of 100 000 iterations of the `emcee` sampler. The red dots represent the mean value of each parameter.

(2018, 3.9 ± 0.4 kpc). However, we do not use the distance to W13 because the *V*-band light curve from Bonanos (2007) is affected by significant non-periodic brightness fluctuations and the eclipses are too shallow to enable accurate modelling the system. We also revisited the detached eclipsing binary $W_{DEB}$ reported by Koumpia & Bonanos (2012), combined their *V*-band light-curve with our $JHK_s$ photometry. A preliminary analysis, however, pointed out that its distance ($d \sim 5$ kpc) is inconsistent with the position of Wd 1 cluster.

### 3.2.4 Confirmation of the W36 A Spectral Type

As a sanity test to check the spectral type of W36 A, we measure the equivalent width (EW) of the He I $\lambda$ 1.7007 $\mu$m and He II $\lambda$ 1.6921 $\mu$m lines from IGRINS/Gemini spectrum. The EW of an absorption line is determined by measuring the width of the adjacent continuum that has the same area as is taken up by the absorption line. Before calculating the EW of W36 A, we have subtracted the secondary component contribution by using the luminosity ratio $L_2/L_1 = 0.32$ (Table 4). Thereafter, two Gaussian profiles were used to fit the absoption lines of both com-ponents (see Fig. 5). The measured EW for He I $\lambda$ 17007 Å and He II $\lambda$ 16921 Å lines were 1.19 ± 0.13 Å and 0.54 ± 0.09 Å, respectively, and log(EW 17007/EW 16921) = 0.34 ± 0.09 dex. We used the correlation between the ratio of the EWs and the spectral type of O-type stars from Lenorzer et al. (2004, see their Fig. 6) to confirm the spectral type of W36 A as an O6.5 III star, in agreement with the K-band classification from Fig. 1. According to Martins et al. (2005), a star of spectral type O6.5 III has an effective temperature of 37 134 K, which agrees with the SPAMMS analysis done at Sect. 3.2.1.

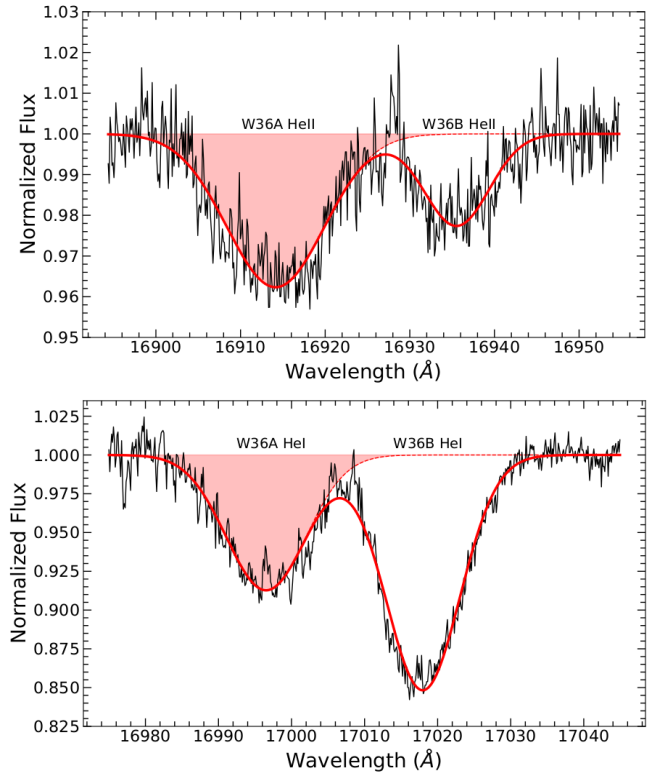

**Figure 5.** Equivalent width of the He I $\lambda$ 17007 Å and He II $\lambda$ 16921 Å lines derived from the IGRINS/Gemini spectrum taken on February 8, 2020. Solid red lines represent the best fit of two Gaussian profiles to the He I $\lambda$ 17007 Å and He II $\lambda$ 16921 Å absorption lines of W36 A and W36 B. Dashed red lines and the red shaded area are the best Gaussian fit and the equivalent width for the absorption lines of W36 A.

### 3.3 Hertzsprung–Russell diagram analysis of Westerlund 1

#### 3.3.1 Age of Stellar sub-Groups

Fig. 6 shows the upper HR diagram. Evolutionary tracks are from Yusof et al. (2022) for supersolar abundances and with rotation ($V/V_{crit}$ = 0.40). Observational data are from Paper I and from Table B1. A number of stars between the extreme ages of the W36 binary (red star) and those of the Red Supergiants (red circles). Several luminous evolved stars are observed with ages 8–11 Myr, including the BSGs (blue diamonds) and YHGs (yellow squares). The most interesting feature is a very populated and compact vertical group around the 7 Myr isochrone (blue diamonds) of Giants and super-Giants.

Given its orbital short period ($P \sim 3.18$ days) the W36 stellar components very likely underwent mass exchange episodes. W36 A as the more luminous should have lost mass, however, it may not have been very large, since its Helium abundance ($n(He)/n(H) = 0.30$) is not much higher than solar. The mass loss of the W36 A had a low impact on its ageing speed, but the peeling effect should have displaced it to the blue in the HR diagram and looks to have a shorter age than in reality. W36 B evolved faster than if it was single and it looks older than in reality, although by a little amount. Although the age of W36 is very uncertain, we can assign it in the broad range 3.5 Myr < $age_{W36}$ < 7.1 Myr.

#### 3.3.2 Discrepancy between evolutionary and dynamic masses

By simultaneously modelling the V-band light curve and the radial velocity curve using WDC and emcee, we estimated the masses for both A and B components of the W36 system as W36A = $22.9^{+2.7}_{-2.9}$ $M_\odot$ and W36B = $16.3^{+2.2}_{-2.3}$ $M_\odot$. The evolutionary masses shown in Fig. 6 are substantially larger (W36A $\sim$ 35 $M_\odot$ and W36B $\sim$ 22 $M_\odot$). The discrepancy between the dynamic and evolutionary masses cannot be explained by mass loss of the stars, for example, otherwise the components would have large Helium overabundance. This problem is common in massive young binaries (Herrero et al. 1992) and the possible causes are: (i) the effect of rotational mixing and tides (e.g., de Mink et al. 2009), and (ii) the underestimation of the convective layer size near the central core by the evolutionary stellar models, as discussed by Tkachenko et al. (2020).

## 4 DISCUSSION AND CONCLUSIONS

We presented a photometric and spectroscopic analysis of the eclipsing binary W36 to derive the distance to the massive young cluster Wd 1. Using the SPAMMS code to disentangle the contributions of each stellar component of the system, together with their radial velocity curves, we were able to limit the parameter space to be fitted by WDC and emcee and characterise the binary system. The physical parameters of the system (see Table 4) were obtained through the simultaneous analysis of the radial velocity and V-band light curves. We derived the masses $M_1 = 22.9^{+2.9}_{-2.5}$ $M_\odot$ and $M_2 = 16.3^{+2.9}_{-2.3}$ $M_\odot$, the radii $R_1 = 12.69^{+0.57}_{-0.54}$ $R_\odot$ and $R_2 = 10.46^{+0.63}_{-0.67}$ $R_\odot$, and the effective temperatures, $T_1 = 37\,100^{+1\,000}_{-1\,100}$ K and $T_2 = 30\,833^{+1\,554}_{-1\,539}$ K.

Following Martins et al. (2005), the luminosities obtained for W36 A and W36 B corresponds to O6.5 III and O9.5 IV, respectively. In particular, the spectral type of W36 A agrees with the classification based on the K-band spectrum (see Sect. 3.1) and with the near-infrared equivalent width analysis (see Sect. 3.2.4), therefore reinforcing our results.

We obtained the distance to the W36 system as $d_{W36}$ = 4.03±0.25 kpc. We used the independent Gaia-EDR3 distance to Wd 1 from Paper I to obtain a weighted mean distance to the cluster as $d_{Wd1}$ = 4.05 ± 0.20 kpc ($m - M$ = 13.04 ± 0.11). The improvement in the distance accuracy from 10 % of the GAIA/parallax method to the 5 % of the eclipsing binary method is in line to what we expected. The very large distance impacts much more the parallax method and in addition the W36 system is well behaved enough to enable robust stellar parameters derivation.

In Paper I, the age of the RSGs was reported as 10.7±1 Myr considering Binary Population and Spectral Synthesis (BPASS) models with solar abundances (Eldridge et al. 2020). Using a grid of stellar models with rotation and supersolar abundances from Yusof et al. (2022), and considering the WR, YHG, RSG and PMS stars of the cluster, we concluded that the star formation in Wd 1 spreads in the range 6 – 12 Myr and happened in more than two episodes. The primary star in W36 puts a lower limit for the cluster age, formally t > 3.5 Myr. Due to the large uncertainties because of mass exchange between the companion stars, we can say that W36 is not older than 6.4 ± 0.5 Myr based on the W36 B maximum age limit.

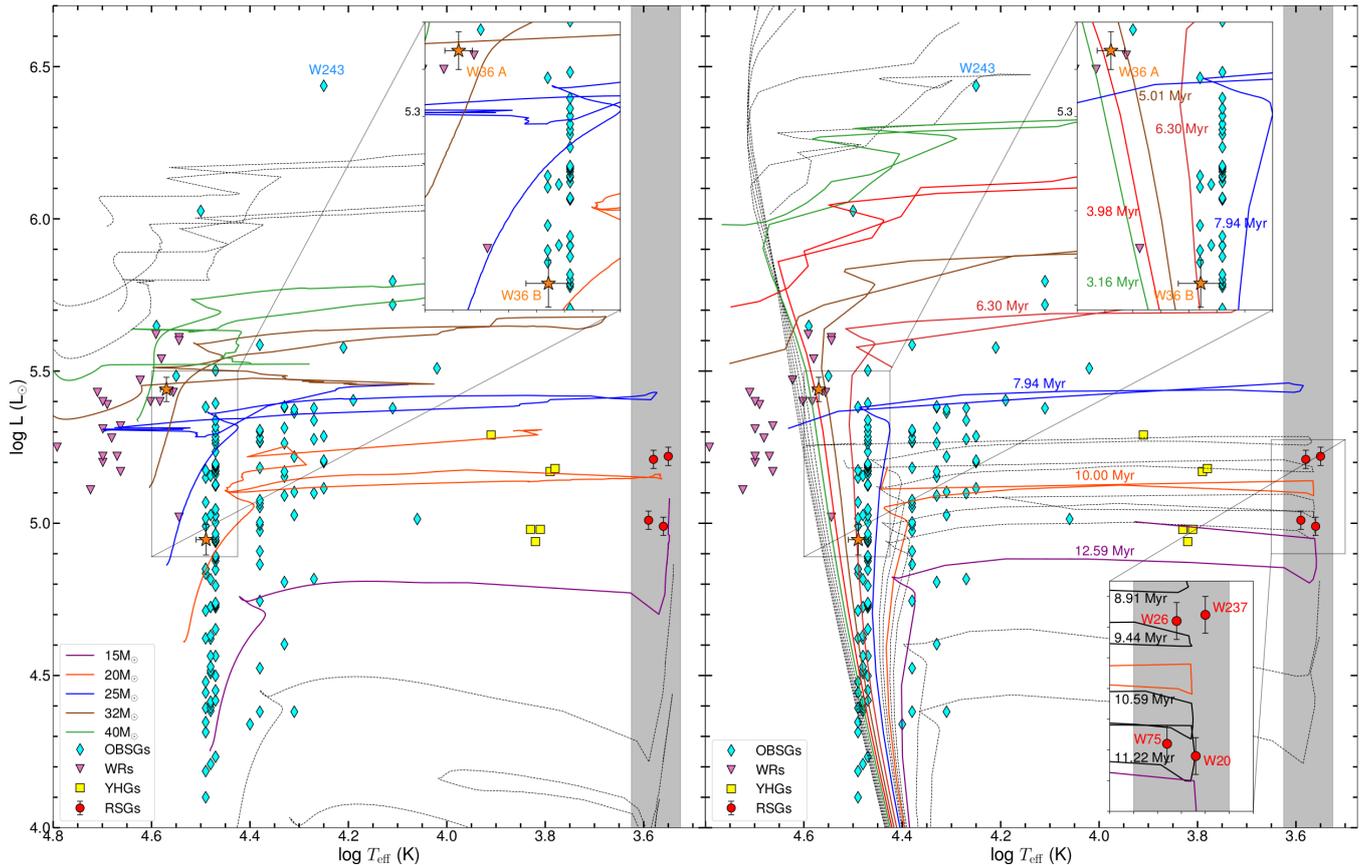

**Figure 6.** HR diagram with mass evolutionary track (left panel) and isochrones (right) of supersolar metallicity (Z=0.020) for rotating models (V/V$_{crit}$ = 0.40) from Yusof et al. (2022). RSGs (red) and YHGs (yellow) are reported in Paper I. OB Giants and Supergiants (blue diamonds) are from Table B1 in the Appendix. Inverted magenta triangles are WRs from Rosslowe (2015), with luminosities scaled to 4.05 kpc. The upper right inset shows W36A and W36B (orange stars) and the one at the bottom, the four RSGs (red circles).

Our *JHKs* photometry revealed other luminous evolved stars with a very conspicuous region of the HRD, containing most of the OB stars, with an age of ∼7 Myr. This group connects nicely with the PMS stars at $7.2^{+1.1}_{-2.3}$ Myr reported by Beasor et al. (2021). Our results confirmed and detailed the scenario proposed by Beasor et al. (2021) and by Yusof et al. (2022) that Wd 1 was not formed in a single starburst event.

He II lines in the *H* and *Ks* bands were critical to derive the correct spectral types of the W36 components, which turned out to be hotter than thought before. This encouraged us to perform high S/N spectroscopy in the NIR region (with the Triplespec/SOAR spectrograph) for all the stars classified in Clark et al. (2020), which will be reported in a future paper.

**ACKNOWLEDGEMENTS**

LAA thanks the Conselho Nacional de Desenvolvimento Científico e Tecnológico (CNPq) for funding through process (315502/2021-5). AD and FN thanks to Fundação de Amparo à Pesquisa do Estado de São Paulo (FAPESP) for support through process number 2019/02029-2 and 2011/51680-6 (AD) and 2017/19181-9 (FN). AD thanks CNPq for support (301490/2019-8). The work of FN is supported by NOIRLab, which is managed by the Association of Universities for Research in Astronomy (AURA) under a cooperative agreement with the National Science Foundation.

Based in part on observations obtained at the Gemini Observatory (processed using the Gemini IRAF package v1.14), which is operated by the Association of Universities for Research in Astronomy, Inc., under a cooperative agreement with the NSF on behalf of the Gemini partnership: the National Science Foundation (United States), National Research Council (Canada), CONICYT (Chile), Ministerio de Ciencia, Tecnología e Innovación Productiva (Argentina), Ministério da Ciência, Tecnologia e Inovação (Brazil), and Korea Astronomy and Space Science Institute (Republic of Korea).

This work used the Immersion Grating Infrared Spectrometer (IGRINS) that was developed under a collaboration between the University of Texas at Austin and the Korea Astronomy and Space Science Institute (KASI) with the financial support of the US National Science Foundation 27 under grants AST-1229522 and AST-1702267, of the University of Texas at Austin, and of the Korean GMT Project of KASI.

Based in part on observations obtained at the Southern Astrophysical Research (SOAR) telescope, which is a joint project of the Ministério da Ciência, Tecnologia e Inovações (MCTI/LNA) do Brasil, the US National Science Foundation's NOIRLab, the University of North Carolina at Chapel Hill (UNC), and Michigan State University (MSU).

Based in part on observations collected at the European Organisation for Astronomical Research in the Southern Hemisphere under ESO programme 091.D-0179, and/or processed data created thereof.

Based in part on observations carried out at Observatório do Pico dos Dias (OPD), which is operated by LNA/MCTI, Brazil.

## DATA AVAILABILITY

The data sets of FLAMES/VLT were downloaded from observations available in the public domain: http://archive.eso.org/.

Table B1 is also available in the CDS.

## APPENDIX A: EMCEE RESULTS FOR THE SIMULTANEOUS FITTING OF THE RADIAL VELOCITY MEASUREMENT FOR W36

## APPENDIX B: TABLE OF WOLF RAYET AND OB STARS IN WESTERLUND 1

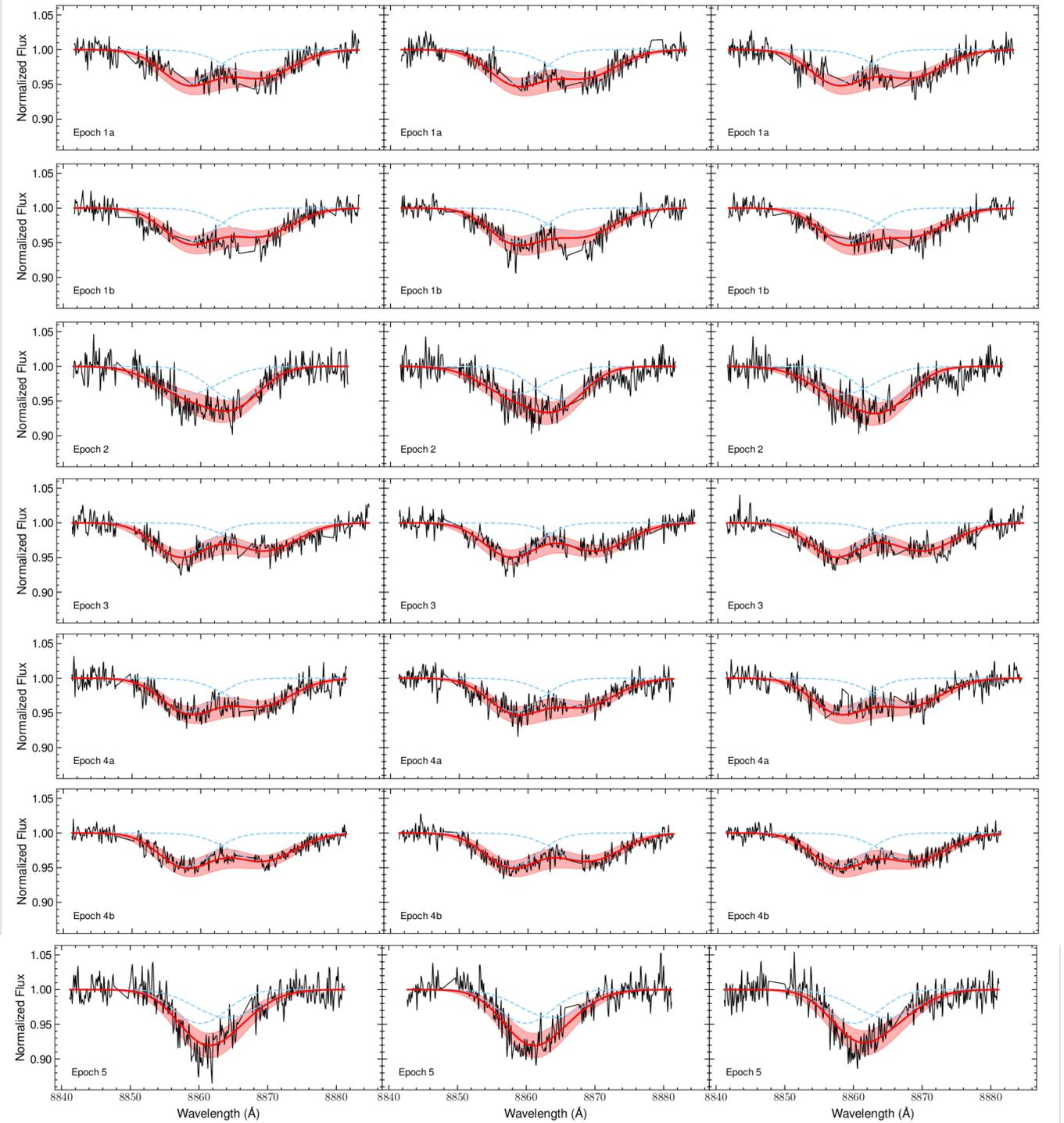

**Figure A1.** Fits of the Pa-11 $\lambda$ 8862 Å spectral lines of the W36 components taken at FLAMES/VLT. Blue-dashed lines represent the individual spectral lines of the primary and secondary stars, red-solid lines are the best fits of the total contribution of both components, and the red-shaded areas representing 1-$\sigma$ around the best fit.

This paper has been typeset from a TeX/LaTeX file prepared by the author.

Table B1. Classification, $JHK_s$ photometry, redenning and bolometric luminosity of the OB stars from Clark et al. (2020). This table is available online at the CDS.

| ID | SpT | $J$ | $H$ | $Ks$ | $AKs$ | $\sigma_{AKs}$ | log $T_{eff}$ | log L (L$_\odot$) |
|---|---|---|---|---|---|---|---|---|
| Wolf Rayet: | | | | | | | | |
| WRA | WN7b+? | - | - | - | - | - | 4.81 | 5.46 |
| WRB | WN7+? | - | - | - | - | - | 4.68 | 5.27 |
| WRC | WC9d | - | - | - | - | - | 4.60 | 5.39 |
| WRD | WN7 | - | - | - | - | - | 4.72 | 5.10 |
| WRE | WC9 | - | - | - | - | - | 4.54 | 5.60 |
| WRF | WC9d | - | - | - | - | - | 4.54 | 5.01 |
| WRG | WN7 | - | - | - | - | - | 4.79 | 5.24 |
| WRH | WC9d | - | - | - | - | - | 4.54 | 5.59 |
| WRI | WN8 | - | - | - | - | - | 4.67 | 5.21 |
| WRJ | WN5 | - | - | - | - | - | 4.58 | 5.39 |
| WRK | WC8 | - | - | - | - | - | 4.70 | 5.30 |
| WRM | WC9d | - | - | - | - | - | 4.69 | 5.38 |
| WRN | WN6o | - | - | - | - | - | 4.59 | 5.61 |
| WRO | WN7 | - | - | - | - | - | 4.62 | 5.46 |
| WRP | WN6 | - | - | - | - | - | 4.66 | 5.31 |
| WRQ | WN5 | - | - | - | - | - | 4.70 | 5.19 |
| WRR | WC9d | - | - | - | - | - | 4.70 | 5.21 |
| WRT | WN8 | - | - | - | - | - | 4.70 | 5.39 |
| WRU | WN6 | - | - | - | - | - | 4.58 | 5.53 |
| WRV | WN5 | - | - | - | - | - | 4.56 | 5.42 |
| WRW | WN7 | - | - | - | - | - | 4.66 | 5.16 |
| WRX | WN5 | - | - | - | - | - | 4.71 | 5.42 |
| OB Supergiants: | | | | | | | | |
| 1001 | O+O? | 12.847 | 12.019 | 11.709 | 0.599 | 0.040 | 4.49 | 4.19 |
| 1002 | O9-9.5II+O? | 12.029 | 11.231 | 10.923 | 0.586 | 0.034 | 4.47 | 4.45 |
| 1003 | O9-9.5bin? | 12.583 | 11.772 | 11.471 | 0.588 | 0.040 | 4.47 | 4.23 |
| 1004 | OeBe | 12.192 | 11.500 | 11.064 | 0.608 | 0.054 | 4.49 | 4.44 |
| 1005 | B0Iab | 10.268 | 9.228 | 8.859 | 0.680 | 0.106 | 4.38 | 5.00 |
| 1006 | O9-9.5IIIbin? | 12.413 | 11.546 | 11.199 | 0.635 | 0.033 | 4.48 | 4.38 |
| 1007 | O9-9.5III | 11.124 | 10.140 | 9.693 | 0.735 | 0.014 | 4.48 | 5.03 |
| 1008 | O9.5II | 11.660 | 10.750 | 10.317 | 0.697 | 0.002 | 4.47 | 4.74 |
| 1009 | B0Ib | 11.441 | 10.496 | 10.054 | 0.679 | 0.048 | 4.38 | 4.52 |
| 1010 | O+O? | 13.226 | 12.327 | 11.976 | 0.650 | 0.039 | 4.49 | 4.10 |
| 1011 | O+O? | 12.440 | 11.567 | 11.226 | 0.634 | 0.037 | 4.49 | 4.39 |
| 1012 | O9-9.5IIIbin? | 11.827 | 10.761 | 10.274 | 0.789 | 0.016 | 4.48 | 4.79 |
| 1014 | O9-9.5IIIbin? | 12.470 | 11.587 | 11.182 | 0.671 | 0.009 | 4.48 | 4.41 |
| 1015 | O9III | 11.442 | 10.495 | 10.122 | 0.681 | 0.040 | 4.49 | 4.85 |
| 1016 | O9-9.5IIIbin? | 12.801 | 11.980 | 11.624 | 0.620 | 0.017 | 4.48 | 4.21 |
| 1017 | O9-9.5IIIbin? | 12.083 | 11.190 | 10.695 | 0.722 | 0.032 | 4.48 | 4.62 |
| 1018 | O9.5Iab | 11.657 | 10.641 | 10.214 | 0.738 | 0.032 | 4.47 | 4.80 |
| 1019 | O9-9.5IIIbin? | 12.517 | 11.667 | 11.220 | 0.679 | 0.020 | 4.48 | 4.39 |
| 1020 | O9-9.5+O? | 11.900 | 11.153 | 10.803 | 0.587 | 0.001 | 4.47 | 4.50 |
| 1022 | O9.5II | 11.873 | 11.065 | 10.684 | 0.628 | 0.001 | 4.47 | 4.56 |
| 1023 | O9III | 11.150 | 10.152 | 9.745 | 0.720 | 0.037 | 4.49 | 5.02 |
| 1024 | O9.5Iab | 11.161 | 10.146 | 9.766 | 0.713 | 0.054 | 4.47 | 4.97 |
| 1025 | O+O? | 12.571 | 11.764 | 11.351 | 0.644 | 0.014 | 4.49 | 4.35 |
| 1026 | O9-9.5III | 12.211 | 11.224 | 10.825 | 0.711 | 0.038 | 4.48 | 4.56 |
| 1027 | O9.5Iab | 11.146 | 9.943 | 9.480 | 0.833 | 0.062 | 4.47 | 5.13 |
| 1028 | O9-9.5IIIbin? | 12.486 | 11.574 | 11.016 | 0.762 | 0.057 | 4.48 | 4.51 |
| 1029 | O9-9.5IIIbin? | 11.973 | 11.000 | 10.480 | 0.768 | 0.023 | 4.48 | 4.73 |
| 1030 | O9.5Ib | 10.311 | 9.463 | 9.115 | 0.627 | 0.027 | 4.47 | 5.19 |
| 1031 | O9III | 11.686 | - | 10.433 | 0.650 | 0.001 | 4.49 | 4.71 |
| 1032 | O9-9.5IIIbin? | 12.482 | 11.606 | 11.173 | 0.683 | 0.006 | 4.48 | 4.41 |
| 1033 | O9-9.5I-III | 11.121 | 10.097 | 9.795 | 0.676 | 0.094 | 4.47 | 4.94 |
| 1034 | O9.5Iab | 11.272 | 10.260 | 9.927 | 0.687 | 0.076 | 4.47 | 4.89 |
| 1035 | O9-9.5IIIbin? | 11.899 | 10.871 | 10.606 | 0.659 | 0.112 | 4.48 | 4.63 |
| 1036 | O9.5Iab | 11.293 | 10.444 | 9.856 | 0.752 | 0.087 | 4.47 | 4.94 |

Notes: (1) Photometric errors are dominated by the zero point calibration to 2MASS ~0.05; (2) Calibration- O-type => Martins et al. (2005). B-type => Wegner (1994) + Koornneef (1983); (3) SP-Type => Clark et al. (2020); (4) BC x $T_{eff}$. Flower (1996); (5) For WRs we used photometry and temperatures from Rosslowe (2015) and his luminosities (at 4 kpc) were scaled to 4.05 kpc. (*) SpT - this work

**Table B1** – *continued*

| ID | SpT | $J$ | $H$ | $Ks$ | $A_{Ks}$ | $\sigma_{AKs}$ | log T$_{eff}$ | log L (L$_\odot$) |
|---|---|---|---|---|---|---|---|---|
| 1037 | O9.5II | 11.526 | 10.550 | 10.085 | 0.741 | 0.004 | 4.47 | 4.85 |
| 1039 | B1Ia | 9.981 | 8.775 | 8.292 | 0.794 | 0.093 | 4.31 | 5.10 |
| 1040 | O9-9.5I-IIIbin? | 10.766 | 9.783 | 9.366 | 0.719 | 0.028 | 4.47 | 5.13 |
| 1041 | O9.5Iabbin? | - | 10.216 | 9.554 | 0.991 | 0.001 | 4.47 | 5.16 |
| 1042 | O9.5II | 11.591 | 10.710 | 10.319 | 0.663 | 0.015 | 4.47 | 4.72 |
| 1044 | O9-9.5IIIbin? | 12.157 | 11.187 | 10.888 | 0.652 | 0.081 | 4.48 | 4.52 |
| 1045 | O9.5II | 11.619 | 10.683 | 10.314 | 0.675 | 0.039 | 4.47 | 4.73 |
| 1046 | O+O? | 11.507 | 10.583 | 10.177 | 0.689 | 0.019 | 4.49 | 4.84 |
| 1047 | O9.5II | 11.311 | 10.411 | 9.944 | 0.710 | 0.017 | 4.47 | 4.89 |
| 1048 | B1.5Ia | 10.448 | 9.244 | 8.745 | 0.802 | 0.085 | 4.27 | 4.82 |
| 1049 | B1-2Ia+ | 8.856 | 7.676 | 7.279 | 0.739 | 0.128 | 4.27 | 5.38 |
| 1050 | O9.5II | 11.674 | 10.767 | 10.329 | 0.698 | 0.001 | 4.47 | 4.73 |
| 1051 | O9III | 10.865 | 9.849 | 9.386 | 0.756 | 0.015 | 4.49 | 5.17 |
| 1052 | O9III | 12.972 | 11.957 | 11.521 | 0.742 | 0.027 | 4.49 | 4.31 |
| 1053 | B0Ib | 10.115 | 8.785 | 8.288 | 0.866 | 0.118 | 4.38 | 5.31 |
| 1054 | O9-9.5bin? | 10.484 | 9.435 | 8.979 | 0.766 | 0.026 | 4.47 | 5.30 |
| 1055 | B0Ib(+O?) | 10.484 | 9.435 | 8.979 | 0.729 | 0.067 | 4.38 | 4.98 |
| 1056 | O9.5II | 11.135 | 10.121 | 9.657 | 0.756 | 0.014 | 4.47 | 5.03 |
| 1057 | O9.5-B0Iab | 10.523 | 9.615 | 9.280 | 0.645 | 0.049 | 4.48 | 5.16 |
| 1058 | O9III | 11.401 | 10.275 | 9.734 | 0.842 | 0.005 | 4.49 | 5.07 |
| 1059 | O9III? | 11.442 | 10.543 | 10.124 | 0.685 | 0.006 | 4.49 | 4.85 |
| 1060 | O9.5II | 11.047 | 10.302 | 9.891 | 0.617 | 0.029 | 4.47 | 4.88 |
| 1061 | O9-9.5IIIbin? | 12.506 | 11.405 | 10.976 | 0.774 | 0.052 | 4.48 | 4.53 |
| 1062 | O+O? | 12.116 | 11.097 | 10.729 | 0.708 | 0.061 | 4.49 | 4.62 |
| 1063 | O9III | 12.648 | 11.535 | 11.133 | 0.765 | 0.068 | 4.49 | 4.48 |
| 1064 | O9.5Iab | 11.857 | 10.651 | 10.022 | 0.920 | 0.017 | 4.47 | 4.95 |
| 1065 | B0Ib | 10.022 | 8.868 | 8.472 | 0.741 | 0.122 | 4.38 | 5.18 |
| 1066 | O9III | 11.282 | 10.219 | 9.827 | 0.739 | 0.060 | 4.49 | 4.99 |
| 1067 | B0Iab | 9.738 | 8.620 | 8.239 | 0.719 | 0.120 | 4.38 | 5.27 |
| 1068 | B0Ib | 10.864 | 9.564 | 8.962 | 0.908 | 0.061 | 4.38 | 5.05 |
| 1069 | B5Ia+ | 8.912 | 7.344 | 6.217 | 0.750 | 0.050 | 4.11 | 5.38 |
| W10 | B0.5 I+OB | 9.362 | 8.382 | 7.733 | 0.791 | 0.040 | 4.33 | 5.38 |
| W11 | B2 Ia | 9.061 | 8.100 | 7.567 | 0.719 | 0.008 | 4.25 | 5.20 |
| W13 | B0.5Ia++OB | 9.144 | 8.153 | 7.662 | 0.714 | 0.038 | 4.33 | 5.38 |
| W15 | O9 Ib | 10.988 | 10.007 | 9.576 | 0.725 | 0.021 | 4.47 | 5.05 |
| W17 | O9Iab | 10.754 | 9.907 | 9.255 | 0.784 | 0.118 | 4.47 | 5.19 |
| W18 | B0.5Ia | 9.405 | 8.495 | 7.952 | 0.707 | 0.007 | 4.33 | 5.26 |
| W19 | B1Ia | 8.993 | 7.956 | 7.537 | 0.692 | 0.081 | 4.31 | 5.36 |
| W21 | B0.5Ia | 9.806 | 8.837 | 8.265 | 0.746 | 0.006 | 4.33 | 5.15 |
| W228b | O9Ib | 11.651 | 10.851 | 10.456 | 0.632 | 0.007 | 4.47 | 4.65 |
| W232 | B0Iab | 10.387 | 9.643 | 9.148 | 0.623 | 0.028 | 4.38 | 4.87 |
| W234 | O9 | 10.412 | 9.643 | 9.148 | 0.670 | 0.063 | 4.47 | 5.20 |
| W238 | B1 Iab | 9.746 | 8.863 | 8.359 | 0.681 | 0.000 | 4.31 | 5.03 |
| W23a | B2 Ia+B I? | 8.952 | 7.951 | 7.387 | 0.751 | 0.003 | 4.25 | 5.29 |
| W24 | O9 Iab | 10.364 | 9.400 | 8.881 | 0.764 | 0.025 | 4.47 | 5.34 |
| w243 | LBV(B2Ia) | 6.246 | 5.271 | 4.570 | 0.811 | 0.069 | 4.25 | 6.44 |
| W25 | O9Iab | 9.866 | 9.129 | 8.451 | 0.752 | 0.158 | 4.47 | 5.50 |
| W27 | O7-8Ia+ | 9.473 | 8.753 | 7.633 | 0.973 | 0.374 | 4.50 | 6.03 |
| W28 | B2 Ia | 8.793 | 7.966 | 7.478 | 0.640 | 0.004 | 4.25 | 5.22 |
| W29 | O9 Ib | 10.621 | 9.694 | 9.143 | 0.765 | 0.050 | 4.47 | 5.23 |
| W2a | B2 Ia | 9.125 | 8.234 | 7.736 | 0.672 | 0.007 | 4.25 | 5.12 |
| W30 | O4-5Ia+ | 10.497 | 9.486 | 8.942 | 0.796 | 0.025 | 4.59 | 5.65 |
| W3002 | B0Iab | 10.354 | 9.370 | 8.905 | 0.707 | 0.046 | 4.38 | 5.00 |
| W3003 | B0Ib | 10.112 | 9.005 | 8.341 | 0.860 | 0.018 | 4.38 | 5.29 |
| W3004 | B0Iab | 9.783 | 8.687 | 8.199 | 0.765 | 0.064 | 4.38 | 5.30 |
| W3005 | O9.5Ib | 10.371 | 9.601 | 9.154 | 0.646 | 0.040 | 4.47 | 5.18 |
| W33 | B5Ia+ | 6.970 | 5.957 | 5.369 | 0.751 | 0.008 | 4.11 | 5.72 |
| W34 | B0 Ia | 9.811 | 8.845 | 8.280 | 0.747 | 0.008 | 4.38 | 5.27 |
| W35 | O9 Iab | 10.432 | 9.617 | 8.950 | 0.778 | 0.134 | 4.47 | 5.32 |
| W36 | O6.5III+O9IV* | 10.506 | 9.327 | 9.091 | 0.690 | 0.060 | 4.55 | 5.48 |
| W37 | O9Ib | 10.581 | 9.600 | 9.219 | 0.699 | 0.045 | 4.47 | 5.17 |
| W373 | B0Iab | 9.733 | 8.833 | 8.585 | 0.560 | 0.129 | 4.38 | 5.07 |
| W38 | O9Iab | 10.904 | 10.052 | 9.342 | 0.816 | 0.145 | 4.47 | 5.17 |

Table B1 – *continued*

| ID | SpT | *J* | *H* | *Ks* | $A_{Ks}$ | $\sigma_{A_{Ks}}$ | log T$_{eff}$ | log L (L$_\odot$) |
|---|---|---|---|---|---|---|---|---|
| W39 | O9Iab | 10.905 | 9.866 | 9.303 | 0.817 | 0.027 | 4.47 | 5.19 |
| W41 | O9Iab | 10.050 | 9.186 | 8.684 | 0.713 | 0.042 | 4.47 | 5.39 |
| w42a | B9Ia+ | 7.231 | 6.041 | 5.343 | 0.868 | 0.016 | 4.02 | 5.51 |
| W43a | B0Ia | 9.099 | 7.909 | 7.500 | 0.759 | 0.123 | 4.38 | 5.62 |
| W43b | B1Ia | 11.312 | 10.472 | 9.961 | 0.658 | 0.012 | 4.31 | 4.38 |
| W43c | O9Ib | 10.735 | 9.810 | 9.131 | 0.830 | 0.112 | 4.47 | 5.26 |
| W46a | B1Ia | 9.293 | 8.195 | 7.619 | 0.798 | 0.022 | 4.31 | 5.37 |
| W46b | O9.5Ib | 10.943 | 9.824 | 9.421 | 0.768 | 0.069 | 4.47 | 5.13 |
| W47 | O9.5Iab | 10.740 | 10.129 | 9.388 | 0.732 | 0.220 | 4.47 | 5.12 |
| W49 | B0Iab | 11.188 | 10.209 | 9.602 | 0.778 | 0.023 | 4.38 | 4.75 |
| W5 | B0.5 Ia+ | 9.678 | 8.807 | 8.201 | 0.723 | 0.047 | 4.33 | 5.17 |
| W50b | O9III | 10.755 | 9.857 | 9.396 | 0.706 | 0.014 | 4.49 | 5.15 |
| W52 | B1.5Ia | 9.052 | 8.078 | 7.528 | 0.733 | 0.003 | 4.27 | 5.28 |
| W53 | OBIa+OBIa | 10.223 | 9.207 | 8.825 | 0.714 | 0.053 | 4.49 | 5.38 |
| W54 | B0.5Iab | 10.062 | 9.085 | 8.442 | 0.786 | 0.038 | 4.33 | 5.09 |
| W55 | B0Ia | 9.987 | 9.016 | 8.659 | 0.642 | 0.092 | 4.38 | 5.08 |
| W56a | B1.5Ia | 9.234 | 8.224 | 7.880 | 0.642 | 0.111 | 4.27 | 5.10 |
| W56b | O9.5Ib | 11.086 | 10.055 | 9.638 | 0.739 | 0.040 | 4.47 | 5.03 |
| W57a | B4Ia | 8.133 | 7.173 | 6.665 | 0.682 | 0.009 | 4.06 | 5.01 |
| W60 | B0Iab | 10.479 | 9.628 | 9.100 | 0.684 | 0.017 | 4.38 | 4.91 |
| W61a | B0.5Ia | 9.256 | 8.424 | 7.824 | 0.704 | 0.054 | 4.33 | 5.31 |
| W61b | O9.5Iab | 10.485 | 9.565 | 9.035 | 0.751 | 0.042 | 4.47 | 5.27 |
| W62a | B0.5Ib | 10.716 | 9.886 | 9.479 | 0.604 | 0.038 | 4.33 | 4.60 |
| W63a | B0Iab | 12.081 | 11.000 | 10.500 | 0.765 | 0.054 | 4.38 | 4.38 |
| W65 | O9Ib | 10.957 | 10.193 | 9.644 | 0.696 | 0.090 | 4.47 | 5.00 |
| W6a | B0.5Iab | 10.265 | 9.600 | 9.000 | 0.635 | 0.096 | 4.33 | 4.81 |
| W6b | O9.5III | 12.638 | 11.665 | 11.173 | 0.753 | 0.010 | 4.47 | 4.42 |
| W7 | B5Ia+ | 6.846 | 5.787 | 5.196 | 0.772 | 0.002 | 4.11 | 5.80 |
| W70 | B3Ia | 8.337 | 7.276 | 6.722 | 0.762 | 0.025 | 4.19 | 5.40 |
| W71 | B2.5Ia | 8.081 | 7.044 | 6.433 | 0.781 | 0.008 | 4.21 | 5.58 |
| W74 | O9.5Iab | 10.360 | 9.367 | 8.968 | 0.714 | 0.039 | 4.47 | 5.28 |
| W78 | B1Ia | 9.300 | 8.396 | 7.890 | 0.681 | 0.006 | 4.31 | 5.25 |
| W84 | O9.5Ib | 10.996 | 10.219 | 9.743 | 0.664 | 0.052 | 4.47 | 4.95 |
| W86 | O9.5Ib | 11.219 | 10.350 | 9.927 | 0.675 | 0.003 | 4.47 | 4.89 |
| W8b | B1.5Ia | 8.762 | 7.620 | 7.317 | 0.675 | 0.163 | 4.27 | 5.34 |